\documentclass{LNCS/llncs}
\usepackage{makeidx}  
\usepackage{epsfig}
\begin{document}
\mainmatter              
\title{Re-visiting the One-Time Pad}
\titlerunning{One-Time Pad}  
\author{Nithin Nagaraj\inst{1}, Vivek Vaidya\inst{2} \and Prabhakar G Vaidya\inst{1}}
\authorrunning{Nithin Nagaraj et al.}   
\tocauthor{Nithin Nagaraj (National Institute of Advanced
Studies), Vivek Vaidya (GE Global Research), Prabhakar G Vaidya
(National Institute of Advanced Studies)} 
\institute{Mathematical Modelling Unit \\ National Institute of Advanced Studies,  IISc Campus, Bangalore, INDIA\\
\email{nithin@nias.iisc.ernet.in, pgvaidya@nias.iisc.ernet.in}\\
WWW home page:
\texttt{http://www.geocities.com/nias\_mmu} \and Imaging Technologies Lab \\
GE Global Research, Bangalore, INDIA\\
\email{vivek.vaidya@ge.com}}

\maketitle              

\begin{abstract}
In 1949, Shannon proved the perfect secrecy of the Vernam
cryptographic system, also popularly known as the One-Time Pad
(OTP). Since then, it has been believed that the perfectly random
and uncompressible OTP which is transmitted needs to have a length
equal to the message length for this result to be true. In this
paper, we prove that the length of the transmitted OTP which
actually contains useful information need not be compromised and
could be less than the message length without sacrificing perfect
secrecy. We also provide a new interpretation for the OTP
encryption by treating the message bits as making True/False
statements about the pad, which we define as a private-object. We
introduce the paradigm of private-object cryptography where
messages are transmitted by verifying statements about a
secret-object. We conclude by suggesting the use of Formal
Axiomatic Systems for investing $N$ bits of secret.
\end{abstract}
\section{Introduction}
Cryptography, the science and the art of communicating messages
secretly has been the subject of intense research for the last 50
years. The field itself is much older, dating back to as old as
1900 BC, when Egyptian scribes, a derived form of the standard
hieroglyphics were used for secure communication.

\par In 1949, Shannon, the father of information theory, wrote a
seminal paper (see \cite{Shannon49}) on the theory of secrecy
systems, where he established the area on a firm footing by using
concepts from his information theory \cite{Shannon48}. In this
lucid paper, among other important contributions, he established
the perfect secrecy of the Vernam cryptographic system, popularly
known as the One-Time Pad or OTP for short. OTP happens to be the
only known {\it perfectly secure} or {\it provably, absolutely
unbreakable} cipher till date. Shannon's work meant that OTPs
offer the best possible mathematical security of any encryption
scheme (under certain conditions), anywhere and anytime $-$ an
astonishing result.

\par There have been a number of other cryptographic algorithms \cite{Menezes96} in the last century, but
none can provide Shannon security (perfect security). This is one
of our motivations to probe into the OTP and investigate its
properties. To the best of our knowledge there has been very
little work on the OTP since Shannon. Recently, Raub and others
\cite{Raub04} describe a statistically secure one time pad based
crypto-system. Dodis and Spencer \cite{Dodis02} show that the
difficulty of finding perfect random sources could make achieving
perfect security for the OTP an impossibility. We shall not deal
with the issue of random sources in this paper. The question we
intend to ask in this paper is $-$ what can we say about the
length of the OTP to be transmitted across the secure channel? We
prove a counter-intuitive result in this paper $-$ the length of
the OTP to be transmitted need not always be equal to the length
of the message and that it is possible to achieve Shannon security
even if the transmitted OTP length is actually smaller than the
message length. Note that we treat the OTP as perfectly random and
uncompressible. However, the length of the OTP is one piece of
information that is not exploited and is always compromised in its
traditional usage. We construct a protocol where this piece of
information can be used effectively to reduce the length of the
OTP to be transmitted while not losing Shannon security for any of
the bits of the message. We then give an alternate interpretation
of the OTP encryption and follow it up with a new paradigm of
cryptography called private-object cryptography.

\par The paper is divided as follows. In the next section, we describe
the OTP and its traditional interpretation as XOR operation by
means of a simple example. In Section~3, we prove the central
theoretical result of the paper $-$ that it is possible to have
the transmitted OTP length less than the message length while
still retaining perfect secrecy. We first prove a 1-bit reduction
of the transmitted OTP length and then generalize for a $k$-bit
reduction for a message of length $N>k$ bits. We also give an
alternative method of compressing the OTP based on the length
information which is universally known. In Section~4, we provide
our new alternate interpretation of the OTP as a private-object
and the encrytpion/decryption as equivalent to making statements
about the object. Section~5 talks more about the new paradigm of
private-object cryptography. We claim that every private-key
cryptography is essentially a form of private-object cryptography
and can provide theoretical security for at least one message of
length equal to the entropy of the crypto-system. We then ask the
important question $-$ how should we invest $N$ bits of secret? We
hint towards the use of Formal Axiomatic Systems (FAS) for this
purpose. We conclude in Section~6.
\section{One-Time Pad}
In 1917, Gilbert Vernam of AT\&T invented the first electrical one
time pad. The Vernam cipher was obtained by combining each
character in the message with a character on a paper tape key.
Captain Joseph Mauborgne (then a Captain in the United States Army
and later chief of the Signal Corps) recognized that the character
on the key tape could be made completely random. Together they
invented the first one time tape system. There were other
developments in the 1920s which resulted in the paper pad system.
The Germans had paper pads with each page containing lines of
random numbers. A page would be used to encrypt data by simple
addition of the message with these random numbers. The recipient
who had a duplicate of the paper pad would reverse the procedure
and then destroy his copy of the page. An OTP was used for
encrypting a teletype hot-line between Washington and Moscow. OTPs
were also used successfully by the English in World War II. These
were especially useful in battlefields and remote regions where
there were no sophisticated equipments for encryption, all that
they used were OTPs printed on silk. The final discovery of the
significance and theoretical importance of the OTP was made by
Claude Shannon in 1949.
\subsection{The Classical Interpretation of OTP}
We describe the encryption and decryption of an OTP by a simple
example. Alice and Bob have shared an OTP ($K=1011001001$) in
complete secrecy (assume that they have met in private and shared
the key). One fine day, Alice wants to invite Bob to her house and
wishes to send the message `COME AT 8 PM' to him. But she is
afraid of the interception of the message by Eve whom she
dislikes. She therefore encrypts her message as follows. She first
converts her message into binary (assume that she has a dictionary
which converts the message into the bits $M=0010110101$). She then
performs an XOR operation to yield the cipher-text $C = K \oplus M
= 1001111100$. She transmits this across a public channel. Bob
receives the cipher-text $C$. Since he has the OTP with him, he
does the XOR operation of the cipher-text with the OTP to yield
the correct message $M = C \oplus K = 0010110101$. He then looks
up at the dictionary (this need not be secret) and converts this
to the more readable message `COME AT 8 PM'
\begin{figure}[h]
\centerline{\includegraphics[width=4.0in]{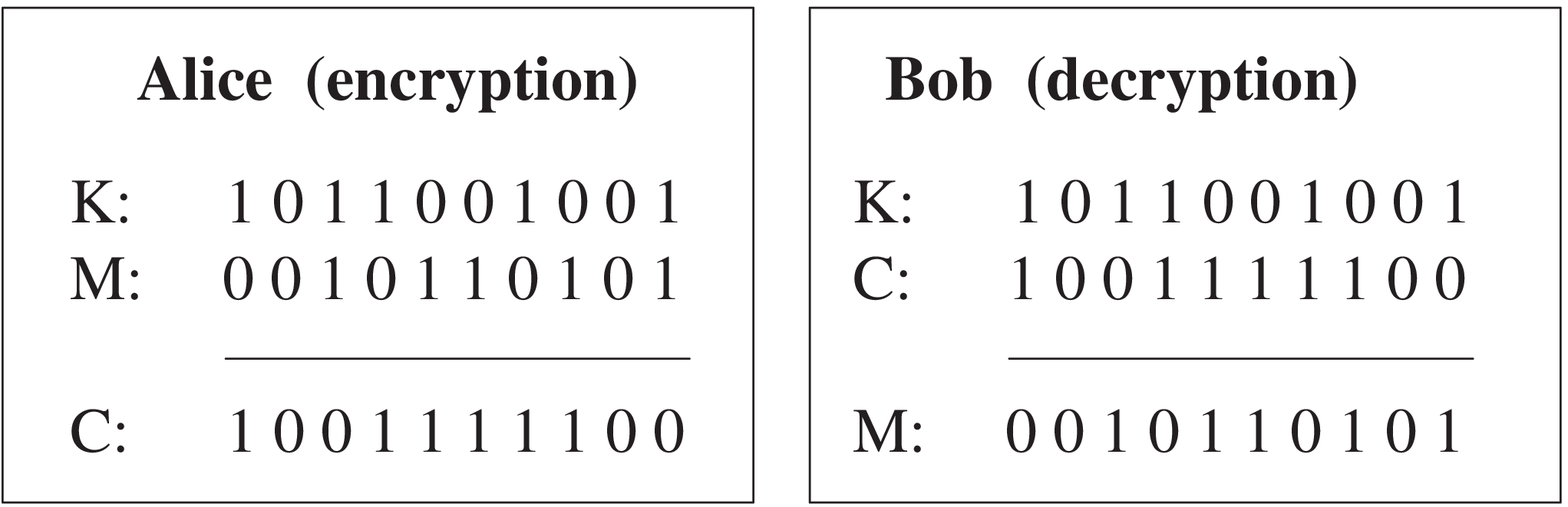}} \caption{
The One-Time Pad encryption and decryption interpreted as XOR
operation} \label{fig:otp}
\end{figure}

To summarize (refer to Fig.~\ref{fig:otp}):
\begin{enumerate}
\item The OTP is a random set of bits which is used as a
private-key known only to Alice and Bob. %
\item The OTP encryption involves an XOR operation of the message
$M$ with the OTP to yield the cipher-text $C$. %
\item The OTP decryption involves an XOR of the cipher-text $C$
with the OTP to get back the original message $M$.%
\end{enumerate}
The classical interpretation of the OTP as XOR implies the
following two important observations.
\begin{enumerate}
\item The length of the OTP is completely compromised in the
process of encryption. %
\item One bit of the OTP is employed to encrypt exactly one bit of
the message and this requires one XOR operation. All bits of the
message require the same amount of effort to encrypt and decrypt.
\end{enumerate}
We shall have more to say about the above observations later. But
what can we say about the security of the OTP encryption?
\subsection{Security of the OTP}
Shannon, in his seminal 1949 paper on the theory of secrecy
systems \cite{Shannon49} defined perfect secrecy as the condition
that the {\it a posteriori} probabilities of all possible messages
are equal to the {\it a priori} probabilities independently of the
number of messages and the number of possible cryptograms. This
means that the cryptanalyst has no information whatsoever by
intercepting the cipher-text because all of her probabilities as
to what the cryptogram contains remain unchanged. He then argued
that there must be at least as many of cryptograms as the messages
since for a given key, there must exist a one-to-one
correspondence between all the messages and some of the
cryptograms. In other words, there is at least one key which
transforms any given message into any of the cryptograms. In
particular, he gave an example of a perfect system with equal
number of cryptograms and messages with a suitable transformation
transforming every message to every cryptogram. He then showed
that the OTP actually achieves this. In other words, the best
possible mathematical security is obtained by the OTP. Incidently,
this is the only known method that achieves Shannon security till
date.
\section{Transmitted OTPs of Length Less than the Message Length}
It has generally been believed that the OTPs that are transmitted
are required to have a length equal to that of the message in
order for Shannon's argument to hold. In this section, we show
this is not the case. Although the length of the OTP {\it while}
encryption need to be equal to the length of the message, the OTP
that is transmitted could be less. But this sounds quite
paradoxical because the OTP is assumed to have been derived from a
perfect random source and hence uncompressible. Even if we are
able to construct a compression algorithm that compresses some of
the generated OTPs, it has to expand some other OTPs, it can't
losslessly compress {\it all} OTPs. This is because of the {\it
Counting Argument} \cite{Salomon00} which states that every
lossless compression algorithm can compress only some messages
while expanding others. However, we prove the central theoretical
result of this paper that the transmitted OTP length can be $0
\leq k < N$ bits less than the message length $N$ while still
retaining perfect secrecy. Although we might not be able to
achieve this reduction all the time, our method {\it never
expands} the transmitted OTP. At worst, our transmitted OTPs are
of the length of the message. We first prove an easier case where
the OTP could be less than the message by 1-bit and the same idea
is employed for the $k-$bit reduction. We make use of our earlier
observation that the OTP encryption compromises its length
in its traditional usage which we can actually avoid.\\

\noindent  {\bf Theorem 1:~} {\it For every message of length $N$
bits, it is equally likely that the transmitted OTP was of length
$N-1$ or $N$ bits while still retaining perfect theoretical secrecy.}\\

\begin{figure}[h]
\centering
\includegraphics[width=4.0in] {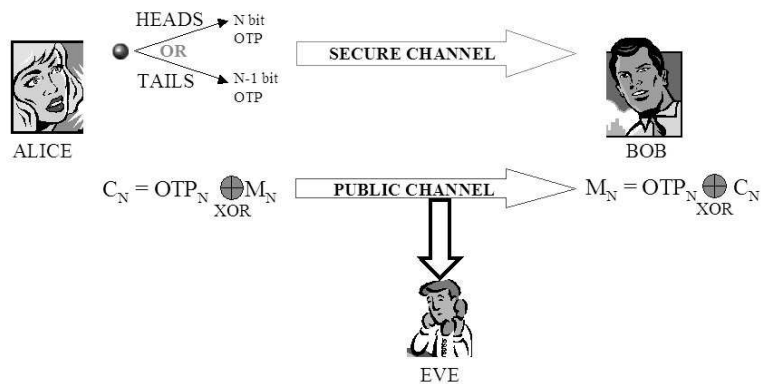} 
 \caption[OTP length reduction] {The protocol for the transmitted OTP length reduction by 1 bit} \label{fig:figotp}
\end{figure}

\noindent  {\bf Proof:~} We shall prove this result by
constructing a (modified) protocol (Fig.~\ref{fig:figotp}) where
Alice and Bob exchange a message of length $N$ by using an OTP.
However, in this modified protocol, there is a 50\% probability
that the transmitted OTP had a length of $N-1$ or $N$ while still
retaining perfect secrecy. We guarantee perfect secrecy for all
the $N$ bits of the message. The protocol
works as follows:\\\\
{\bf Step 1:~} Alice performs a coin flip with a perfect coin. If
it falls {\bf HEADS}, she constructs an OTP of length $N$ and if
it falls {\bf TAILS} she constructs an OTP of length $N-1$. It is
assumed that Alice has access to a perfect random source to
construct the OTP in either
events.\\\\
{\bf Step 2:~} Alice communicates the OTP through a secure channel
to Bob. \\\\
{\bf Step 3:~} On some later day, Alice intends to send a message
of length $N$ bits to Bob. If the OTP she generated has $N-1$
bits, she appends an additional bit at the end of the OTP. This
additional bit is set to $1$ if the length $N-1$ is {\bf ODD} and
to $0$ if $N-1$ is {\bf EVEN}. In case the OTP already has $N$
bits, Alice forces the $N^{th}$ bit to $0$ if $N-1$ is {\bf ODD} and to $1$ if $N-1$ is {\bf EVEN}.\\\\
{\bf Step 4:~} Alice then performs the XOR operation of the
message with the resulting OTP to yield a cipher-text
$C$ which has $N$ bits. She transmits $C$ on the insecure public channel to Bob. \\\\
{\bf Step 5:~} Bob receives $C$. Bob checks to see if the OTP he
had earlier received from Alice has sufficient bits to decrypt the
message. In other words, does it have $N$ bits or $N-1$ bits. In
case the OTP has $N-1$ bits, he does the exact same trick which
Alice did i.e. appends an additional bit and sets it to $1$ or $0$
depending on whether $N-1$ is {\bf ODD} or {\bf EVEN}
respectively.
If the OTP already has $N$ bits, Bob forces the $N^{th}$ bit to $0$ if $N-1$ is {\bf ODD} and to $1$ if $N-1$ is {\bf EVEN}.\\\\\
{\bf Step 6:~} Bob decrypts $C$ by performing an XOR with the
modified OTP and obtains the message.\\

We need not prove the perfect secrecy of the first $N-1$ bits as
Shannon's arguments hold. We need to prove that the $N^{th}$ bit
is perfectly secure. We shall analyze the situation from the
eavesdropper Eve's perspective. Eve knows of this entire protocol.
Eve intercepts the cipher-text $C$ which is of length $N$ bits.
She knows that there is a 50\% probability that it came from an
OTP which originally had $N-1$ bits or $N$ bits. She has no other
strategy but to make a random guess and the probability of success
is 50\%. Hence, her guess of the $N^{th}$ bit is no better than a
50\% success. This proves the perfect secrecy of the $N^{th}$ bit.

While this result seems highly theoretical and of little practical
value, it actually shows an interesting aspect of the OTP which
has been taken for granted. The fact that the length of the OTP
contains {\it information} is usually neglected. Our proof was
aimed at achieving theoretical security for one additional bit by
using the Least Significant Bit (LSB) of the length of the OTP (by
{\bf ODD} we mean LSB$=1$ and {\bf EVEN} we mean LSB$=0$) and we
can do this half of the time. The natural question to ask is $-$
can we make use of the other bits of the length?
\subsection{$k$ Bit Reduction in the Length of the Transmitted OTP}
\noindent  {\bf Theorem 2:~} {\it For every message of length $N$,
it is possible that the transmitted OTP had one of the lengths $N
- k, N - k + 1, \ldots, N-2, N-1$ or $N$ with respective
probabilities $2^{-k}, 2^{-k}, \ldots, 2^{-k}$ or $1-k2^{-k}$
while still retaining perfect theoretical secrecy.}\\

\noindent {\bf Proof:~} We generalize the aforementioned argument
for a $k-$bit reduction in the length of the transmitted OTP.
Assume that the message length $N>k$ and let the binary
representation of the numbers $N, N-1, N-2, \ldots, N-k$ be the
following:~$<A0_{m_N} \ldots A0_{2}A0_{1}>, <A1_{m_{N-1}} \ldots
A1_{2}A1_{1}>, <A2_{m_{N-2}} \ldots A2_{2}A2_{1}>, \ldots,
<Ak_{m_{N-k}} \ldots Ak_{2}Ak_{1}>$ where each of the $Ai_j$ is
binary for all $i$ and $j$. Also $m_{N} = \lfloor log_2N \rfloor +
1, m_{N-1} = \lfloor log_2(N-1) \rfloor + 1, \ldots, m_{N-k} =
\lfloor log_2(N-k) \rfloor + 1$ (note that $m_{N-k} \geq 1$ since
$N
> k$). Alice has a $k$-sided biased coin which produces OTPs of
length $N, N-1, N-2, \ldots, N-k$ with probabilities $1-k2^{-k},
2^{-k}, 2^{-k}, \ldots, 2^{-k}$. For encryption of a $N$ bit
message, Alice forces the last $k$ bits of the $N-1$, $N-2$,
$\ldots$, $N-k$ OTPs to the bits $<A1_{N-k} \ldots A1_{2}A1_{1}>,
<A2_{N-k} \ldots A2_{2}A2_{1}>, \ldots, <Ak_{N-k} \ldots
Ak_{2}Ak_{1}>$ respectively. Only for the instance when Alice is
generating an OTP of length $N$ bits, she ensures that the last
$N-k$ bits never have the same sequence as the other $k$ OTPs
before sending it to Bob on the secure channel. Moreover, she
ensures that the remaining available combinations which are $2^k -
k$ in number have each a probability of occurrence $=2^{-k}$. This
way, the last $k$ bits of the OTPs are perfectly random because
the probability of obtaining any particular sequence of $k$ bits
is $2^{-k}$. The rest of the protocol remains unchanged.

\par With this, we have proved by construction that it is possible for
the transmitted OTP to have a length lesser than the message
length with a non-zero probability while still attaining perfect
theoretical secrecy. It is interesting to see that for larger
reductions (larger values of $k$), the probability of obtaining a
reduction reduces. The best average reduction is for $k=2$ and
$k=3$, where we get $0.75$ bits of reduction. The average
reduction is given by $k(k+1)2^{-(k+1)}$. The upper bound on $k$
is given by the condition $m_{N-k} \geq k$ which implies $N \geq k
+ 2^{(k-1)}$. Note that in our protocol, we have not violated the
assumption that the OTP is perfectly random and incompressible.
\subsection{Compression of Transmitted OTP Based on Length Information}
Alternatively, we can say that the transmitted OTP is compressible
to the extent the length information allows. We provide a method
of compressing the transmitted OTP given the fact that the
messages to be encrypted are always of length $N$, which is
publicly known.

\par Alice generates an $N$ bit OTP. If the last bit is 1, she deletes
it to create an $N-1$ bit OTP. If the last bit is 0, she deletes
all bits which are zeros from the end up to and including the bit
which is 1. If the OTP has no 1s in it, then Alice transmit it as
is. As an example, consider the $N=10$ bit OTP `1011001001'. Since
the last bit is 1, Alice deletes to create the 9-bit OTP
`101100100'. If the $N=10$ bit OTP happens to be `1011001000', by
the above rule, Alice obtains the 6-bit OTP -- `101100'. Alice
transmits the resulting {\it compressed} OTP across the secure
channel to Bob. Since the length of messages to be encrypted is
always $N=10$, Bob decompresses the received OTP to $N$ bits by
reversing the rule. In other words, if Bob receives a $N-1$ bit
OTP, he appends a 1 to make it $N$ bits. If the received OTP is of
length $N-k$ bits, where $k>1$, he appends a 1 followed by $k-1$
zeros. Thus, the OTP is correctly decompressed by Bob in all
instances.

\par An interesting thing to observe is that the OTP is compressed for
all instances except the case when it has no 1s. There is only one
such OTP (all 0s) which is uncompressed by this scheme. At a first
glance, one might wrongly infer that we are contradicting the
counting argument. However, this is not the case. The counting
argument applies only to {\it memoryless} lossless compression
algorithms. In our case, Bob has the {\it a priori} information
about the length $N$ and hence it is not memoryless.

\par What are the reductions obtained by this method? We can see that
for 50\% of the instances, there is a reduction by 1-bit only (the
last bit is 1 for 50\% of the cases). Among the remaining 50\%,
one instance is uncompressed (the OTP with all bits 0s) and one
instance has a maximum reduction of all $N$ bits (the OTP with a 1
followed by $N-1$ bits). For the remaining OTPs, the compression
ratios vary depending on the number of 0s in the end. For example,
an OTP with $m$ zeros in the end has a reduction of $m+1$ bits.
There are $2^{N-m-1}$ such $N-$bit OTPs which will compress to an
OTP of length $N-m-1$ bits, a reduction by $m+1$ bits.
\section{An Alternate Interpretation of the OTP as a Private-Object}
In the previous section, we saw how we made use of the length of
the OTP in obtaining a reduction in its length. The length happens
to be a particular feature of the OTP, as if it were an {\it
object}. This leads us to the notion of a private-object which we
define as follows.\\

\noindent {\bf Private-Object:~} Any object which is known only to
the
sender and the receiver is defined as a {\it private-object}.\\

\par The above definition is very broad. The object may have any
embodiment, not necessarily digital in nature. The object could be
a real physical thing or it could be an one time pad (could even
be multi-dimensional). An important thing to note is that every
private-object enables theoretically secure communication. The
entropy of the private-object is determined by the {\it number of
independent True/False statements} that can be made about the
object without revealing any information about it. The way a
message is transmitted by means of a private-object is described
below.

\par Alice and Bob share a private-object $P$, known only to them.
Alice intends to send a message $M$ (as an example, the statement
`COME AT 8 PM' to Bob). The protocol is as
follows:\\

\noindent {\bf Step 1:~} Alice converts message $M$ into binary
representation (using a publicly known dictionary). Say `COME AT 8
PM' translates to $M = 0010110101$.\\

\noindent {\bf Step 2:~} Alice substitutes $0=TRUE=T$ and
$1=FALSE=F$.
Therefore $M=TTFTFFTFTF$.\\

\noindent {\bf Step 3:~} For each bit of the message $M$, Alice
makes statements about the private-object $P$ which is TRUE (if
the bit is T) or FALSE (if the bit is F) to obtain the cipher-text
$C$. In other words $C =
<statement_1><statement_2>\ldots<statement_{10}>$ where
$<statement_1>$ is TRUE, $<statement_2>$ is TRUE, $<statement_3>$
is FALSE etc. As a crude example, assume that the private-object
is a physical object which has 3 eyes, 2 hands, 5 legs etc. Alice
could make a statement like `$P$ has 3 eyes' which is TRUE or a
statement like `$P$ has 4 legs' which is FALSE (the number of legs
and hands in this hypothetical object are independent of each other).\\

\noindent {\bf Step 4:~} Bob receives the cipher-text $C$ which is
a collection of statements about $P$. He verifies each statement
and determines whether they are TRUE ($T$) or FALSE ($F$). He
obtains
a string of $T$s and $F$s by this process ($M=TTFTFFTFTF$).\\

\noindent {\bf Step 5:~} Bob substitutes $0=TRUE=T$ and
$1=FALSE=F$ in $M$
to obtain the binary message $M = 0010110101$.\\

\noindent {\bf Step 6:~} Bob looks up at the dictionary for $M$ to
obtain
the message `COME AT 8 PM'.\\

\par The OTP can be thought of as a private-object $P$ and the above
protocol can be used for secure communication. For our previous
example of Section 2, the set of statements which Alice would make
are $C =$ `the first bit of the OTP is 1', `the second bit of the
OTP is 0' $\ldots$ `the tenth bit of the OTP is 0'. Bob verifies
these statements since he has the OTP with him and obtains the
correct message.
\section{Private-Object Cryptography}
In the previous section, we saw how the OTP could be viewed as a
private-object and statements about the object can be made to
transmit information securely. So long as the statements are {\it
independent} of each other, we are guaranteed to achieve perfect
secrecy. This is because every statement encrypts one bit of the
message and is making use of a unique feature of the
private-object. For the OTP, every bit is its unique independent
feature. For private-objects of the real physical world, the
features could be the number of edges or the number of faces etc.
Determining the number of {\it unique} and {\it independent}
features in a physical object might be difficult. This means that
the {\it entropy} of the object is difficult to compute. The
amount of information that can be securely transmitted by this
method is upper bounded by the entropy of the object in bits.
Private-key or symmetric-key cryptography is a subset of
Private-object cryptography where the key happens to be a set of
bits on which various mathematical operations are made. In effect,
every private-key crypto-system is only making statements about
the key which is the private-object. Since the key of a
private-key is usually much shorter than the message, the
statements are not {\it independent} of each other. They formally
map to complex statements about the key. We state the
following theorem without proof:\\

\noindent {\bf Theorem 3:~} {\it Every symmetric-key crypto-system
can encrypt exactly one binary message having a length equal to
the
entropy of the crypto-system with perfect theoretical secrecy.} \\

\par One can always make a certain number of {\it unique} and {\it
independent} statements about the crypto-system. We can treat the
crypto-system with its unique parameters as a private-object
having a certain entropy. These statements are {\it finite} in
number and can be used to communicate a finite length binary
message with perfect secrecy (equivalent to an OTP of the same
entropy). The length of the message can be at most equal to the
entropy of the crypto-system without sacrificing Shannon security.
Finding the entropy of the crypto-system may not always be easy.

\par Another interesting off-shoot is the definition of the {\it
entropy of an  object} of the real world. We can invert our above
observation to say that the entropy of an object is the number of
bits of information that can be transmitted with perfect secrecy
by making independent statements about the object. In other words,
we claim that there exists a mapping from every object of the real
world to an OTP and the entropy of that OTP is the entropy of the
object. It may be hard in practice to determine the entropy of
objects.
\subsection{Investment of N-bits of Secret}
Let us now relax the {\it perfect secrecy} constraint since we
need to send long keys (if not as long as the message) for
achieving this. Assume that we have a fixed bit-budget, say $N$
bits of secret. We wish to know what is the best private-object to
invest these $N$ bits of secret so as to achieve a high encryption
efficiency. Here, we do not wish to achieve perfect secrecy, but
breaking the system should be very hard. Here, we are being vague
in our definition. It suffices to say that we wish to obtain a
method where currently known methods of cryptanalysis have a hard
time in breaking, if not impossible. We wish to propose using a
Formal Axiomatic System (FAS) for investing these $N$ bits. This
part of the paper is a bit speculative in nature and is mainly a
motivation towards potential future research.
\subsection{FACtS: Formal Axiomatic Cryptographic System}
A Formal Axiomatic System or FAS for short, refers to a system of
axioms and rules of inference which together define a set of
theorems \cite{Hofstadter99}. An example of a FAS is Typographical
Number Theory (TNT). Hilbert's program was to completely formalize
the whole of Mathematics using TNT. This ambitious plan was
derailed by G\"{o}del who proved that all consistent and
sufficiently powerful axiomatic systems contain {\it undecidable}
propositions. Because of this, Formal Axiomatic Systems are
fascinating objects.

\par
\begin{figure}[h]
\centering
\includegraphics[width=2.5in] {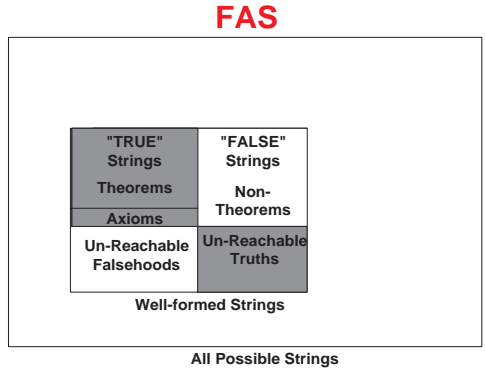} 
 \caption[String Space of a FAS] {The string-space of a Formal Axiomatic System (FAS)} \label{fig:figfas}
\end{figure}

 We can view a FAS in another interesting way which is the {\it
compression} view-point. A FAS is actually a {\it compressed}
version of all its theorems which can be proven within the system.
It is this viewpoint that motivates us to consider an FAS as a
private-object which is shared between Alice and Bob. If Alice
were given a bit-budget of $N$ bits, she could invest it in the
construction of a FAS which is {\it consistent} and {\it
sufficiently strong}. These are the only two requirements. She
would have to define a set of axioms and rules of inference to
completely specify the FAS. She shares this as a {\it
private-object} with Bob over a secure channel. The way Alice and
Bob can now exchange information is to make statements or strings
in the FAS. The receiver can {\it verify} whether a particular
statement or string is TRUE or FALSE in the FAS which they share.
If it is TRUE, then it implies that the string is a {\it Theorem}
and the bit conveyed is $0$. If the string is FALSE, then it is a
{\it Non-theorem} and conveys the bit $1$. We basically use the
private-object paradigm with Theorems and Non-theorems of the FAS
acting as binary representations for $0$ and $1$ respectively. We
name such a system as Formal Axiomatic Cryptographic System
(FACtS). Fig.~\ref{fig:figfas} shows the string space of a FAS
\cite{Hofstadter99}.

\par Since the FAS is sufficiently strong, it would contain
G\"{o}delian statements which are undecidable (the system is
incomplete). We believe that it may be possible to {\it confuse}
and {\it diffuse} the cryptanalyst by a clever use of G\"{o}delian
statements in the cipher-text. This is a speculation on our part,
because we do not know of any procedure which would enable us to
construct such statements in large numbers.

 \par One of the biggest advantages of such a set-up is the difficulty
of breaking the system for Eve using brute-force attack. In
conventional systems such as the RSA and other public-key and
private-key methods \cite{Menezes96}, brute-force attack would
involve trying out all possible keys in the {\it key-space}. For
example, if the key length is 128-bit, it would mean trying out
$2^{128}$ (a huge number) guesses for the private-key. A computer
could mechanically try out these number of possibilities until it
found the right key. This would probably take a long time but with
a number of computers in parallel or by using Quantum computers,
this time could be sufficiently reduced. The important thing to
realize in this scenario is that there is a {\it mechanical
procedure} for trying out all the combination and with exponential
increase in computational power over time, it could be eventual
broken (eg: the RSA-128 is already broken). In our system, the
equivalent would be to try out all possible Formal Axiomatic
Systems of a given length $N$. However there would be several
systems which are {\it duds}, those that are inconsistent or
meaningless. Computers which are designed to try out different
FASs might have a difficult time to find out inconsistencies. They
might have to deal with the Turing Halting problem.
\section{Conclusions}
To summarize, the central contribution of this paper is a new
result in the OTP literature. We have shown that the length of the
OTP which is traditionally compromised in encryption could be
avoided. We proved that it is possible to reduce the key-length of
the transmitted OTP (which is perfectly random and uncompressible
otherwise) while still retaining perfect secrecy. Even though this
reduction is small, it is nevertheless useful in saving band-width
for crypto-systems which use OTPs on a regular basis (we showed
that we never expand the OTPs in any case unlike compression
algorithms which always expand some). We have conceived a new
paradigm called private-object cryptography which makes use of
statements about an object (private to the communicating parties)
for secure message transmission and showed how the OTP can be
re-interpreted in this new paradigm. We also claimed that all
existing private-key crypto-systems are a form of private-object
cryptography. Further, they are in essence making statements about
the secret key. We believe that these statements are not
independent but are necessarily more complex. We then suggested
the investment of $N$ bits of secret in a FAS. The verification of
strings or statements of the FAS as theorems or non-theorems could
convey a bit of information. It may be the case that the structure
of the FAS and the space of theorems and non-theorems could be
designed so that it is {\it sufficiently random} for cryptographic
purposes. More research needs to be done in these directions.
%
%


\begin{thebibliography}{5}
%
\bibitem {Shannon49}
Shannon, C.: Communication Theory of Secrecy Systems. Bell System
Technical Journal {\bf 28} (1949) 656--715

\bibitem {Shannon48}
Shannon, C.: A Mathematical Theory of Communication. Bell System
Technical Journal {\bf 27} (1948) 379--423


\bibitem {Menezes96}
Menezes, A., van Oorschot, P.C., Vanstone, S.: Handbook of Applied
Cryptography.  CRC Press, Boca Raton, Florida. (1996)

\bibitem{Raub04}
Raub, D., Steinwandt, R., Mueller-Quade, J.: On the Security and
Composability of the One Time Pad. Cryptology ePrint Archive.
Report 2004/113 (2004) http://eprint.iacr.org/2004/113/.

\bibitem {Dodis02}
Dodis, Y., Spencer, J.: On the (non)Universality of the One-Time
Pad. Proceedings of the 43rd Symposium on Foundations of Computer
Science. (2002) 376--388

\bibitem {Salomon00}
Salomon, D.: Data Compression:The Complete Reference. 2nd edn.
Springer-Verlag, New York Inc. (2000)

\bibitem {Hofstadter99}
Hofstadter, D.: G\"{o}del, Escher, Bach: An Eternal Golden Braid.
20th Anniv. edn. Basic Books (1999)
\end{thebibliography}
\end{document}